\RequirePackage{snapshot}
\documentclass{JHEP3}
\usepackage{relsize,balance,url,xspace,amsmath,textcomp,bm,cite}
\usepackage{hepnicenames,abhepexpt,abmath}
\usepackage{underscore}

\makeatletter
\xspaceaddexceptions{\cite}
\let\@oldcite\cite
\renewcommand{\cite}[1]{~\!\!\@oldcite{#1}\xspace}
\renewcommand{\textrightarrow}{to\xspace}
\newcommand{\etal}{\textit{et~al.}\xspace}
\newcommand{\@Rplus}{\protect\nolinebreak{\smaller\ensuremath{\mspace{-0.5mu}\raisebox{0.3ex}{+}}}}
\newcommand{\@plusplus}{\@Rplus\@Rplus}
\newcommand{\CC}{C\@Rplus\@Rplus\xspace}
\newcommand{\CCHerwig}{Herwig\@Rplus\@Rplus\xspace}
\newcommand{\Fortran}{\textsc{Fortran}\xspace}
\newcommand{\Sherpa}{\textsc{Sherpa}\xspace}
\newcommand{\Perl}{Perl\xspace}
\newcommand{\Python}{Python\xspace}
\newcommand{\Java}{Java\xspace}
\newcommand{\CEDAR}{CEDAR\xspace}
\newcommand{\Rivet}{Rivet\xspace}
\newcommand{\RivetGun}{RivetGun\xspace}
\newcommand{\HZTool}{HZTool\xspace}
\newcommand{\HZSteer}{HZSteer\xspace}
\newcommand{\JetWeb}{JetWeb\xspace}
\newcommand{\HepData}{HepData\xspace}
\newcommand{\HepForge}{HepForge\xspace}

\newcommand{\HepML}{HepML\xspace}
\newcommand{\HepMC}{HepMC\xspace}
\renewcommand{\vec}[1]{\bm{#1}}
\newcommand{\kbd}[1]{\texttt{#1}}
\makeatother

\author{%
Andy Buckley\\
Institute for Particle Physics Phenomenology, Durham University, UK.\\
E-mail: \email{andy.buckley@durham.ac.uk}
}
\title{\CEDAR: tools for event generator tuning}

\preprint{}

\abstract{%
  I describe the work of the \CEDAR collaboration in developing tools for tuning
  and validating Monte Carlo event generator programs. The core \CEDAR task is
  to interface the Durham \HepData database of experimental measurements to
  event generator validation tools such as the UCL \JetWeb system --- this has
  necessitated the migration of \HepData to a new relational database system and
  a Java-based interaction model.
  The ``number crunching'' part of JetWeb is also being upgraded, from the
  \Fortran \HZTool library to the new \CC \Rivet system and a generator
  interfacing layer named \RivetGun.
  Finally, I describe how \Rivet is already being used as a central part 
  of a new generator tuning system, and summarise two other \CEDAR activities, 
  \HepML and \HepForge.}

\begin{document}

\section{Introduction}
Monte Carlo event generators are an essential tool for particle physics,
simulating aspects of collider events ranging from the parton-level signal
process to cascades of QCD and QED radiation in both initial and final states,
non-perturbative hadronisation, underlying event physics and specific particle
decays. Event generators provide experimentalists and phenomenologists with
samples of fully exclusive events drawn from physical distributions, and are
therefore central to the design of both detector hardware and data analysis
strategies: this is more true than ever for LHC physics.

However, event generators are not fully predictive: various phenomenological
parameters must be tuned to experimental data to bootstrap a general purpose
generator before physically meaningful predictions can be obtained. Such
parameters include the parton density functions (PDFs) of the colliding beam
particles, parton shower cutoffs and evolution variables, the running of
$\alpha_s$, choice of $\Lambda_\text{QCD}$, and a variety of hadronisation
parameters which strongly depend on the hadronisation model being applied.
Observable distributions calculated from a generator's output depend on these
parameters in an extremely non-trivial way. This leads to the generator tuning
problem --- how to choose the parameter sets that give the best fits to
experimental data, given that we have no ready parameterisation of the output,
and an exhaustive exploration of the many-dimensional parameter space is out of
the question?

In this paper, I will describe the work done by the \CEDAR
collaboration\cite{cedar:chep04,cedar:web} towards addressing this problem for
the LHC. The main theme of this work is the development of tools to generate and
efficiently analyse simulated events, and the validation of generator tunings
against experimental data using these tools.

\section{\HepData}
\HepData\cite{oldhepdata:web} is a database of experimental particle physics
data, which has been maintained at Durham University since the 1970s. \HepData's
contents are not the raw experimental data, but the data as presented in the
plots and data tables of peer-reviewed experimental papers. \HepData contains
data from a wide range of collider and fixed target experiments, covering many
initial states and $\sqrt{s}$: it is therefore an excellent reference point for
the distributions of observables that event generators should be tuned against.

\subsection{The legacy database}
Since its inception, \HepData has been based on a hierarchical database
management system (HDBMS). In the intervening years, the database world has
firmly centred its attention on more flexible relational database systems
(RDBMS), and today a wide range of high quality RDBMS systems are available for
free, with strong support for the SQL query language and networked availability.
By contrast, the hierarchical systems have evolved little or not at all, and
even simple operations like query changes or schema updates are substantial
tasks involving writing \Fortran routines.

These shortcomings of the legacy database meant that, with the incentive of
placing \HepData as a data service at the centre of projects like \CEDAR, the
decision was taken to upgrade \HepData to use a relational database
system\cite{hepdatajetweb:chep06,hepdata:web}.

\subsection{The new database}
\HepData's data model is intrinsically hierarchical: a given data point is
located within the hierarchy $\text{\textsl{paper}} \rightarrow
\text{\textsl{dataset}} \rightarrow \text{\textsl{axis}} \rightarrow
\text{\textsl{point}}$ and there is little point in comparing data points from
different distributions. Fortunately, a relational model is flexible enough that
implementing a hierarchical structure is easy.

In practice the RDBMS used is MySQL\cite{mysql}, but due to the
ANSI-standardised SQL query language this is easily swappable for almost any
other modern RDBMS. A major design principle has been the decoupling of database
access and implementation details, such as which parts of a data record are
stored in which fields and tables, from the semantic model of how various
aspects of the data are related to each other. Hence, the semantics of the data
are reflected in a \Java ``object model'', which makes no reference to database
implementation, and the Hibernate\cite{hibernate} object-RDBMS persistency
system is used as a layer to isolates ``client'' code from the database
implementation details, using \Java mechanisms such as reflection and
annotations. With Hibernate, the persistency between the persistent database
storage and the ephemeral in-memory object representation is governed by
configuration files and \Java metadata, and so much tedious and fragile database
access coding is avoided.

On top of this object model, the primary mode of access to the database will
remain the Web interface. This is also written in terms of \Java objects,
executed in the Apache Tomcat\cite{tomcat} servlet engine.  More decoupling is
desirable here, and we intend to use the Tapestry\cite{tapestry} templating
system to isolate the data content from the Web presentation according to the
model-view-controller (MVC) design idiom, but as yet there has been little
effort expended on the new Web interface.

While the Web interface, when completed, will be the primary method of searching
and browsing the database, our design means that the object model can also be
imported from, and exported to, various other representations. Datasets will be
representable as graphics (in a variety of formats), plain text, the
AIDA\cite{aida} XML and ROOT data formats and \HepML\cite{hepml:web}. The last
of these is the canonical file format representation of \HepData records and
mechanisms for data import and export in \HepML format are built into \HepData,
using the Castor\cite{castor} object-XML marshalling system. Another form of
\HepML is used by \JetWeb and will be mentioned later.

\subsection{Database migration}
Migrating from the legacy hierarchical database to a new relational database
has proven to be a substantial task.  In large part the difficulties have been
due to the relative unstructured form of the legacy database, which has no
strong type system: data entries are simply text strings. This means that the
structure and integrity of the data has been subject to the whims of various
data submitters over a long time period, tempered by the diligence of the
database managers, and there are many cases where a data record which appears
attractively formatted when presented as a Web page turns out to be hard to fit
into a more rigidly defined data model. To attempt to decouple the target
relational database design from the vagaries of the legacy system, the migration
of data has become a multi-stage process.

\paragraph{Legacy DBMS \textrightarrow{} flat files}
The first stage is to use a mixture of \Fortran and \Perl scripts to massage the
hierarchical data records into a set of tab-delimited plain text files, each of
which contains all the information for one aspect of all the papers in the
database. As might be expected, the data point values and errors files are
extremely large. This procedure needs only be done infrequently, if the legacy
database changes substantially, and so an immediate decoupling is achieved.

\paragraph{Flat files \textrightarrow{} \HepML}
The second stage is to transform these ``flat files'' into a series of \HepML
files, one for each paper, using a \Python script. In practice, we have found
that the efficiency of the \HepML builder can be greatly increased by splitting
each of the flat files so that there is one file per legacy paper, and then
sorting the content of these files. This reduces the number of scans required
through large files, and allows caching to be made use of during the building of
the XML object model. There is not an exact match between the papers from the
legacy database and those which are available at the end of the \HepML-building
procedure because, due to technical limitations, long papers often had to be
split into several parts in the legacy system: the ``logical'' papers which
result from the \HepML builder are the preferable form.

\paragraph{\HepML \textrightarrow{} RDBMS}
The final stage of the \HepData migration is the reading of \HepML files into
the database. While the previous two migration steps are only of use in the
migration process, importing records from \HepML is a core feature of the
\HepData system. The mechanism used is an object persistency framework, Castor,
of which we are only using the XML part. Like Hibernate, Castor describes how
the object model will be stored in a persistent form, but in this case the
persistency medium is an XML string rather than database tables. As some
features of the \HepML representation are not strictly hierarchical, the
JDOM\cite{jdom} \Java XML processing library is used as an input filter when
importing records from \HepML files.

\section{\JetWeb}
The second main component of \CEDAR is \JetWeb\cite{jetweb} an application
for validating the performance of various generator tunings. \JetWeb combines a
system for running a variety of event generator programs, a database of
distributions calculated from simulated events and a Web interface written in
\Java and run on a Tomcat \Java server. \JetWeb's development was motivated by
the need to avoid misleading tunings, where one distribution is fitted at the
expense of unseen others: accordingly, tunings considered by \JetWeb will be
compared to as many distributions as possible. 

Consistently generating data for such a large number of distributions requires
both a good understanding of the physics models involved and a lot of
computational power: \JetWeb helps here by providing a relatively user-friendly
way of configuring the models, by archiving generated data in such a way that
extra statistics can be requested through the Web interface, and providing a
browsable archive of stored results. \JetWeb shows the overall $\chi^2$ for a
chosen model against all distributions, as well as the fit quality to individual
plots, so the overall quality of a given tuning can be readily assessed.

\JetWeb was initially developed at UCL, based on analyses using the
\HZTool\cite{hztool,heralhc05,hztool:web} library and various versions of the
Herwig\cite{herwig65} and Pythia\cite{pythia} \Fortran event generators.  The reference data in this version
was transcribed from a variety of sources, including \HepData. Extension of
\JetWeb's prototype generator interface to deal with generators other than
Pythia and Herwig proved difficult, and hence \CEDAR's work on \JetWeb has
centred on improving the way that generators are modelled, adding mechanisms for
combining generator runs, and separating run parameters from model
parameters\cite{hepdatajetweb:chep06}. The Web user interface has also been
considerably enhanced, and a version of \HepML for describing generator and run
configurations has been developed and incorporated into the \JetWeb system. The
other way in which \JetWeb's modernisation shows is that event generation now
uses Grid resources and authentication rather than local batch farms.

Another \JetWeb change planned under \CEDAR is the linking of \JetWeb to
\HepData --- a ``single-sourcing'' approach which should result in more robust
data--MC comparisons and easier adding of new analyses.  Since the \HepData
migration process has been more lengthy than anticipated, it is only recently
that \JetWeb has begun reading some data directly from the \HepData database
(via the \HepData \Java object model). Before \JetWeb's internal database of
experimental reference data can be eliminated, some extra datasets must be added
to \HepData and the \HepData migration must be essentially complete so that data
entries are reliably retrievable.

\JetWeb is a relatively sophisticated tool for steering event generators based
on ``high level'' event type requests, and for comparing generated data to
reference plots, but it does not actually interface to the various generator
codes or analyse the generated events directly. These r\^oles are filled by a
pair of native applications for event analysis and steering --- in the existing
incarnations of \JetWeb the \Fortran programs \HZTool\cite{hztool} and
\HZSteer\cite{hzsteer} are used, but \CEDAR has developed \CC replacements for
these, titled \Rivet\cite{hztoolrivet:chep06,rivet:web} and
\RivetGun\cite{rivetgun:web} respectively.

\section{\Rivet and \RivetGun}
\Rivet is a \CC replacement for the \Fortran \HZTool library, initially
developed for the HERA experiments \emph{H}1 and \emph{Z}EUS (hence the ``HZ'').
HZTool as a library has two r\^oles: firstly, to provide a collection of physics
utility routines for calculating commonly used quantities, such as
implementations of jet definitions; and second, to collect a set of analyses
which use these routines and produce histograms comparable with experimental
results. As a result of its HERA legacy, the majority of \HZTool analyses are
from DIS and photoproduction experiments.

Initially, \HZTool analyses included code to specifically configure particular
generators. However, this approach scales badly, and a concerted effort was made
in 2005 to decouple \HZTool routines from generator specifics. The result was a
steering package, \HZSteer\cite{hzsteer}, which contains (almost) all of the
generator-specific code, and the current version of \HZTool is purely concerned
with the physics analyses of event records, and not where any given event came
from. The current version of \JetWeb uses \HZTool and \HZSteer to generate and
analyse events.

Even as \HZSteer was being split off from \HZTool, it was clear that time was
running short for \Fortran-based analysis systems. The rising prominence of
\CC-based generators, such as \CCHerwig\cite{herwig++:v2,herwig++:web},
\Sherpa\cite{sherpa} and Pythia 7/8\cite{thepeg:2004,pythia8}, was evident, and
\Fortran does not have the level of sophistication as an application framework
language to steer these generators\footnote{Indeed, technical concerns with how
  \CC encodes symbol names mean that steering \CC from anything other than \CC
  is troublesome.}. A not unimportant secondary point is that the success of a
system like \HZTool relies on the support of the community in providing new
analyses to keep pace with the appearance of new data: the \textit{de facto}
language used by LHC-era experimentalists is \CC, and an analysis system written
in any other language is less likely to be embraced. The result is a new,
\CC-based analysis system, \Rivet, to replace \HZTool, and a generator steering
system, \RivetGun, to replace \HZSteer.

\subsection{\Rivet}
Rivet\cite{rivet:web,hztoolrivet:chep06} (an acronym for ``Robust Independent
Validation of Experiment and Theory'') is a generator-agnostic analysis
framework. Guiding design principles include the implementation in object
oriented \CC and compatibility with existing standard data formats such as
\HepMC\cite{hepmc} and AIDA\cite{aida}. \Rivet has no dependence on
generator-specific features and sees data only via \HepMC event records either
supplied from file or a generator steering package such as \RivetGun (see
below). This makes it easier to incorporate new Monte Carlo generators into a
\Rivet-based validation system than is currently the case with \HZTool.

The Rivet analysis system is based on a concept of "event projections", which
project a simulated event into a lower-dimensional quantity such as scalar or
tensor event shape variables. Projections can be nested and their results are
automatically cached to eliminate duplicate computations, using \CC runtime type
information (RTTI) and comparison operators between projection classes. The
infrastructure has been designed to place as little burden as possible on the
authors of projection and analysis classes, which should be concerned almost
entirely with the analysis algorithm.

Sets of standard projections and analyses are included with the Rivet package,
and this collection will grow with subsequent releases. Analysis data is
accumulated using the AIDA interfaces, and exported primarily in the AIDA XML
histogram format. If ROOT is present on the build system, ROOT format files can
also be exported, allowing use of \Rivet for n-tuple based analyses as well as
the primary design purpose of semi-automated event generator validation. To
complement the generated analysis data, HepData-generated AIDA records for each
bundled analysis are included in the Rivet package and can be used to define the
binnings of generated data observables: this improves the robustness of analysis
implementations and allows easy data-theory comparisons without requiring
network access to HepData.

At the time of writing, the stable version of \Rivet is 0.9, available from the
\Rivet development website\cite{rivet:web}. This first version includes 5
analyses --- two from Tevatron Run 2, one from LEP and two from HERA --- as well
as the library of projections which currently includes $\Ppositron\Pelectron$
event shapes, DIS kinematical boosts, the \DZero ``improved legacy cone'' and
$k_\perp$ jet algorithms via KtJet\cite{ktjet,ktjet:web}, a variety of final
state projections including particle vetoing, and several others. With the main
\Rivet design now stable, we intend for the next release to have much more
substantial libraries of both analyses and projections, such that \Rivet can
entirely replace \HZTool.

\subsection{\RivetGun}
\Rivet is primarily a code library for use by generator steering packages,
although it also includes a command line tool, \kbd{rivet}, which can read in
\HepMC ``ASCII'' event files. The main tool for running \Rivet is the \RivetGun
generator steering program. \RivetGun is written in \CC and provides a uniform
programmatic and command line interface to running event generators, with common
generator configuration features such as setting initial states, named control
parameters and random seeds possible through the most general level of the
interface. Using runtime dynamic library loading, even different versions of the
same generator can be used from the same executable, which would not be possible
with compile time library linking.

The \RivetGun \CC class structure uses inheritance to define a base class,
\kbd{Generator} on which these operations are declared --- each specific
generator then implements the methods declared in the interface in its own way,
be that by mapping common blocks, calling \Fortran library routines or using the
\CC class methods of the target generator. With this approach, switching between
generators with the same initial state configuration is trivial, although
knowledge of generator-specific parameters is still needed for any proper study.
Formally, the generator interfaces are part of a library called AGILe (``A
Generator Interface Library''), since we wish to keep open the possibility of
using the interface without using \Rivet at all.

\RivetGun currently provides generator interfaces for \Fortran
Herwig\cite{herwig65} and Pythia\cite{pythia}, plus enhanced versions of those
generators using the AlpGen\cite{alpgen}, Charybdis\cite{charybdis} and
Jimmy\cite{jimmy,jimmy:web} auxiliary generators.  Preliminary bindings to the
\CCHerwig\cite{herwig++:v2,herwig++:web}, \Sherpa\cite{sherpa,sherpa:web} and
Pythia 8\cite{pythia8} generators are also available. \RivetGun has not yet been
formally released, but the development code is sufficiently usable that it is by
far the easiest way to generate distributions using \Rivet.

Since \RivetGun is intended to be run from within \JetWeb, the generator
configuration will eventually be able to be set from \HepML generator
description files, as well as the current methods of command line arguments
and/or simple key--value parameter files.

\section{Tuning vs. validation}
So far we have addressed efforts towards \CEDAR's central goal, which is the
\emph{validation} of existing event generator tunings. An obvious criticism of
this is that nowhere in the framework is there a procedure for finding a better
or, ideally, optimal tuning. Let us consider the general problem before
describing a particular, \CEDAR-centric solution.

\subsection{The tuning problem}
Na\"ively, one might expect that an event generator can be optimally tuned by
either grid-scanning the parameter space, evaluating a goodness of fit (GoF)
measure against reference data and choosing the best point. Anyone with
experience in sampling problems will be aware of the fallacies at work here: at
the root of the difficulties is the exponential scaling, $\ofOrder{A^n}$, of
computational requirements with the dimensionality $n$ of the parameter space.
This makes comprehensive scanning unrealistic for typical tuning problems with
$n \isOfOrder{10}$. Even adaptive grid-scanning, where the grid is non-uniform,
or adapts to the local GoF, is subject to the exponential scaling.

Sampling specialists may suggest a Markov chain Monte Carlo (MCMC) approach to
this problem. MC sampling works because the scaling is independent of $n$,
although there are only rules of thumb available for estimating sampler
convergence rates and the choice of proposal distribution and use of gradient
information can have very significant effects. However, the typical burn-in for
an MCMC sampler is likely to be in the hundreds or thousands of iterations even
with a well-chosen MCMC setup: this is fine for easily evaluated functions, but
the ``function'' we are attempting to minimise (the GoF for, say 100,000 events
generated with the proposed parameter vector) renders this approach on the edge
of current computational feasibility.

There is a second reason to be wary of such approaches, though: a global GoF
measure is not sensitive enough for a non-exhaustive search such as MCMC to find
a global optimum. There are simply too many ways for parts of distributions to
fit better or worse than others and the combinatorics generate a parameter space
vastly dominated by mediocre tunings --- what is needed is a method which is
more sensitive to the dependence of each element of each distribution on the
tuning parameters. Such a method was used by the LEP \Delphi
experiment\cite{delphi:tune,hamacher}, and it is this approach which we will now
describe.

\subsection{Professor --- tuning with bin-by-bin interpolation}
The \Delphi approach to event generator tuning was to fit a function to the
generator output on a bin-by-bin basis and then to minimise the goodness of fit
in all bins simultaneously, using the interpolating functions. While previous
approaches fitted a linear function, \Delphi fitted a second order polynomial
since that is the first order at which inter-parameter correlations are taken
account of\cite{hamacher}.

\Delphi's \Fortran code implementing this algorithm was called Professor, and
so is its continuation, although the implementation language is now \Python
combined with the \CC \Rivet and \RivetGun systems. Professor is not a \CEDAR
project --- it is a collaborative effort between the Durham IPPP and TU Dresden
--- but its aims are so closely connected to \CEDAR that it seems prudent to
mention it here.

The procedure implemented by Professor is as follows:
\begin{enumerate}
\item Define a hypercube in the $n$-dimensional parameter space, by specifying
  sample ranges in each parameter.
\item Generate $N$ random parameter $n$-vectors in the hypercube. There is no
  upper limit on the number of samples --- indeed, the more the better as might
  be expected, --- but there is a minimum number, given later.
\item Run \RivetGun and \Rivet using the sampled parameter values: this will
  produce $N$ \Rivet output AIDA files, each of which, say, describes $B$ bins.
  The number of events generated in each run should be sufficient to reduce
  statistical error to a near-negligible level.
\item For each bin, $b$, fit a polynomial function to the $N$ generated values,
  using a singular value decomposition (SVD)\cite{svd}. The SVD allows
  calculation of a ``pseudoinverse'' --- a matrix inverse for non-square
  matrices\cite{moore:pseudoinverse,penrose:pseudoinverse} whose use in
  overconstrained systems performs a least-squares fit\cite{penrose:bestsoln}.
  The function to be fitted is the general second-order polynomial in $n$
  dimensions,
  \begin{align}
  f_b(\vec{p}) = 
    \alpha_{b} + 
    \sum_i \beta_{b,i} \, \vec{p'}_i + 
    \sum_{i,\,j \ge i} \gamma_{b,ij} \, \vec{p'}_i \, \vec{p'}_j .
  \end{align}
  where $\vec{p'} = \vec{p} - \vec{p}_0$, the shift of the parameters from their
  nominal/central values $\vec{p}_0$ and $\alpha_b$, $\beta_{b,i}$ \&
  $\gamma_{b,ij}$ are the sets of polynomial coefficients to be determined for
  each bin $b$. Coefficient counting reveals that there must be $N \ge 1 + (n^2
  + 3n)/2$ parameter space samples for there to be an inverse. In principle,
  $\vec{p}_0$ itself can be considered as a set of parameters to be fitted,
  which would add another $n$ to the minimum $N$.
\item Use reference data from \HepData to compute a goodness of fit function for
  each bin, such as error-weighted square deviation, $\phi_b(\vec{p}) =
  (f_b(\vec{p}) - r_b)^2/E_b^2$ where $r_b$ and $E_b$ are the experimental value and
  uncertainty respectively.
\item Analytically or numerically minimise the individual $\phi_b$ functions and
  the corresponding global GoF figure, the ubiquitous $\chi^2$,
  defined\footnote{We're being somewhat sloppy about the definition of the error
    in this $\chi^2$ definition: strictly it should be the ``theory error'' to
    avoid a biased distribution.} as $\chi^2(\vec{p}) = \sum_b \phi_b(\vec{p})$.
  Flag any significant deviations of the bin-by-bin minima from the global
  minimum.
\item Generate a final MC event sample using the interpolation-optimal
  parameters and compare with the prediction.
\end{enumerate}

Professor is currently in active development, testing the method with toy models
and low-dimensional samplings using \Rivet and \RivetGun. The first uses of it
will be on relatively simple cases such as re-implementing the original \Delphi
optimisation, and then proceeding to more complex tunings for the LHC where data
from various generators must be combined and extrapolated. There is a great
opportunity for statistical sophistication in this area, including bin--bin
correlations, various GoF measures\cite{stattoolkit:gof} and weighting of
particular distributions and bins.



\section{\HepForge}
As a spin-off from our own development requirements, \CEDAR now provides a free
online collaborative development facility,
\HepForge\cite{hepforge:chep06,hepforge:web}, for HEP projects which aim to
provide useful, well-engineered tools to the community.

\HepForge currently offers feature-enhanced Web hosting, hosting and HTTP access
to the Subversion code management system (a modern replacement for CVS), an
integrated bug tracker and wiki system strongly integrated with Subversion and
mailing lists for developer contact, project announcements and discussion. All
the \CEDAR projects and about 20 others are hosted with \HepForge and it has
proven a popular alternative to CERN's Savannah system, particularly for small
phenomenology collaborations.

The long-term plan for \HepForge is that it will provide search facilities for a
wide range of HEP computational tools, but at present we are consolidating our
developer support and improving the existing system.

\section{Summary}
\CEDAR is providing a wide variety of computational tools, which are the wide
foundation on which a systematic and global validation and tuning of HEP event
generators is to be built. The largest-scale components of \CEDAR are \HepData
and \JetWeb --- established resources which have been substantially re-designed
and upgraded by \CEDAR. \HepData's migration from a legacy hierarchical database
to a much more rigorously structured relational database and \Java object model
and persistency system has been a substantial task and is now approaching its
final stages. \JetWeb has been internally restructured a great deal, but the
most obvious consequence of the \CEDAR upgrades is the forthcoming use of
\HepData as a source of reference data. The \HepML XML formats provide the glue
between these systems, and a file persistency format.

At a finer-grained level, the \Rivet and \RivetGun systems are the \CC
replacements for \HZTool and \HZSteer being created by \CEDAR. The first
official release of \Rivet has recently taken place and work is now proceeding
on adding new analyses to it and preparing \RivetGun for its first stable
release. At \Rivet's core is the concept of event projections and generator
independence --- with these and a design aim to make it as easy as possible to
write algorithm-focused analysis code, \Rivet is an excellent framework for LHC
validation and tuning studies and users are encouraged to try it out.

Finally, mention was made of the Professor event generator tuning effort and of
the \HepForge development environment. These will respectively build on the
foundation provided by \CEDAR and continue to provide facilities for the
development of HEP computational tools.

\section{Acknowledgements}
The \CEDAR team would like to thank the UK Science and Technology Funding
Council (STFC) for their generous support. \CEDAR's work has also been supported
in part by a Marie Curie Research Training Network of the European Community's
Sixth Framework Programme under contract number \mbox{MRTN-CT-2006-035606}.


\begin{thebibliography}{99}
\bibitem{cedar:chep04}{\mbox{J.\,M.~Butterworth}~\etal, \mbox{hep-ph/0412139}, presented at CHEP'04, Interlaken, September 2004}
\bibitem{cedar:web}{\CEDAR Web site: \url{http://www.cedar.ac.uk}}
\bibitem{oldhepdata:web}{Legacy \HepData Web interface: \url{http://durpdg.dur.ac.uk/hepdata/}}
\bibitem{hepdatajetweb:chep06}{\mbox{A.~Buckley~\etal}, ``HepData and JetWeb'' in {\it Proceedings of CHEP'06}, arXiv:hep-ph/0605048v1}
\bibitem{hepdata:web}{\HepData development: \url{http://projects.hepforge.org/hepdata/}}
\bibitem{mysql}{MySQL: \url{http://www.mysql.com}}
\bibitem{hibernate}{Hibernate: \url{http://www.hibernate.org}}
\bibitem{tomcat}{Apache Tomcat: \url{http://tomcat.apache.org}}
\bibitem{tapestry}{Tapestry: \url{http://tapestry.apache.org}}
\bibitem{aida}{Abstract Interfaces for Data Analysis (AIDA), \url{http://aida.freehep.org/}}
\bibitem{hepml:web}{\HepML: \url{http://projects.hepforge.org/hepml/}}
\bibitem{castor}{Castor: \url{http://www.castor.org}}
\bibitem{jdom}{JDOM: \url{http://www.jdom.org}}
\bibitem{jetweb}{\mbox{J.\,M.~Butterworth and S.~Butterworth}, \mbox{Comput.~Phys.~Commun.~{\bf 153}~(2003)~164}, \url{http://jetweb.cedar.ac.uk/}}
\bibitem{hztool}{\mbox{J.~Bromley~\etal}, ``HZTOOL: A package for Monte Carlo-data comparison at HERA (version~1.0),'' in {\it Workshop on Future Physics at HERA, 1996}}
\bibitem{heralhc05}{\mbox{J.\,M.~Butterworth, H.~Jung, V.~Lendermann, B.\,M.~Waugh}, in {\it HERA and the LHC: A Workshop on the implications of HERA for LHC physics}, Proceedings,~Part~B., hep-ph/0601013}
\bibitem{hztool:web}{\HZTool: \url{http://hepforge.cedar.ac.uk/hztool/}}
\bibitem{herwig65}{\mbox{G.~Corcella~\etal}, ``HERWIG 6.5 release note,'' arXiv:hep-ph/0210213}
\bibitem{pythia}{\mbox{T.~Sj\"ostrand, S.~Mrenna and P.~Skands}, \mbox{JHEP~{\bf 0605}~(2006)~026}, arXiv:hep-ph/0603175}
\bibitem{hzsteer}{\HZSteer: \url{http://hepforge.cedar.ac.uk/hzsteer/}}
\bibitem{hztoolrivet:chep06}{\mbox{B.\,M.~Waugh~\etal}, ``HZTool and Rivet'' in {\it Proceedings of CHEP'06}, arXiv:hep-ph/0605034v1}
\bibitem{rivet:web}{\Rivet: \url{http://projects.hepforge.org/rivet/}}
\bibitem{rivetgun:web}{\RivetGun: \url{http://projects.hepforge.org/rivetgun/}}
\bibitem{hepmc}{\mbox{M.~Dobbs and J.\,B.~Hansen}, \mbox{Comput.~Phys.~Commun.~{\bf 134}~(2001)~41}, \url{https://savannah.cern.ch/projects/hepmc/}}
\bibitem{herwig++:v2}{\mbox{S.~Gieseke~\etal}, ``\CCHerwig 2.0 release note,'', arXiv:hep-ph/0609306}
\bibitem{herwig++:web}{\CCHerwig: \url{http://projects.hepforge.org/herwig/}}
\bibitem{sherpa}{\mbox{T.~Gleisberg, S.~Hoche, F.~Krauss, A.~Schalicke, S.~Schumann and J.~C.~Winter}, ``SHERPA 1.alpha, a proof-of-concept version,'', \mbox{JHEP~{\bf 0402}~(2004)~056}, arXiv:hep-ph/0311263}
\bibitem{sherpa:web}{\url{http://projects.hepforge.org/sherpa/}}
\bibitem{thepeg:2004}{L.~L\"onnblad, ``ThePEG, PYTHIA7 and ARIADNE,'', prepared~for {\it 12th International Workshop on Deep Inelastic Scattering (DIS 2004)}}
\bibitem{pythia8}{Pythia 8: \url{http://www.thep.lu.se/~torbjorn/pythiaaux/future.html}}
\bibitem{alpgen}{\mbox{M.~L.~Mangano, M.~Moretti, F.~Piccinini, R.~Pittau and A.~D.~Polosa}, \mbox{JHEP~{\bf 0307}~(2003)~001}, arXiv:hep-ph/0206293}
\bibitem{charybdis}{\mbox{C.~M.~Harris, P.~Richardson and B.~R.~Webber}, \mbox{JHEP~{\bf 0308}~(2003)~033}, arXiv:hep-ph/0307305}
\bibitem{jimmy}{\mbox{J.\,M.~Butterworth, J.\,R.~Forshaw and M.\,H.~Seymour}, ``Multiparton interactions in photoproduction at HERA,'', \mbox{Z.~Phys.~C~{\bf 72}~(1996)~637}, arXiv:hep-ph/9601371}
\bibitem{jimmy:web}{Jimmy: \url{http://hepforge.cedar.ac.uk/jimmy/}}
\bibitem{ktjet}{\mbox{J.\,M.~Butterworth, J.\,P.~Couchman, B.\,E.~Cox and B.\,M.~Waugh}, ``KtJet: A \CC implementation of the $k_\perp$ clustering algorithm,'' \mbox{Comput.~Phys.~Commun.~{\bf 153}~(2003)~85}, arXiv:hep-ph/0210022}
\bibitem{ktjet:web}{KtJet: \url{http://hepforge.cedar.ac.uk/ktjet/}}
\bibitem{delphi:tune}{\mbox{P.~Abreu~\etal}, \mbox{DELPHI Collaboration}, \mbox{Z.~Phys.~C~{\bf 73}~(1996)~11.}}
\bibitem{hamacher}{\mbox{K.~Hamacher and M.~Weierstall}, ``The Next Round of Hadronic Generator Tuning Heavily Based on Identified Particle Data,'' arXiv:hep-ex/9511011}
\bibitem{svd}{H.~Abdi, {\it Singular Value Decomposition (SVD) and Generalized Singular Value Decomposition (GSVD)}, in {N.\,J.~Salkind~(Ed.)}, {\it Encyclopedia of Measurement and Statistics}, Thousand Oaks, 2007}
\bibitem{moore:pseudoinverse}{E.\,H.~Moore, ``On the reciprocal of the general algebraic matrix'', in {\it Bulletin of the American Mathematical Society} \mbox{{\bf 26}~(1920)~394--395}}
\bibitem{penrose:pseudoinverse}{R.~Penrose, ``A generalized inverse for matrices'', in {\it Proceedings of the Cambridge Philosophical Society} \mbox{{\bf51}~(1955)~406--413}}
\bibitem{penrose:bestsoln}{R.~Penrose, ``On best approximate solution of linear matrix equations'', in {\it Proceedings of the Cambridge Philosophical Society} \mbox{{\bf 52}~(1956)~17--19}} 
\bibitem{stattoolkit:gof}{\mbox{S.~Donadio, S.~Guatelli, B.~Mascialino, M.~G.~Pia, A.~Pfeiffer, A.~Ribon and P.~Viarengo}, ``A statistical toolkit for data analysis,'', \mbox{Nucl.~Phys.~Proc.~Suppl.~{\bf 150}~(2006)~50}}
\bibitem{hepforge:chep06}{\mbox{A.~Buckley~\etal}, ``HepForge: A lightweight development environment for HEP software'' in {\it Proceedings of CHEP'06}, arXiv:hep-ph/0605048v1}
\bibitem{hepforge:web}{\HepForge: \url{http://www.hepforge.org}}
\end{thebibliography}
\end{document}